\documentclass[twocolumn,showpacs,aps,prl,superscriptaddress,nofootinbib,floatfix]{revtex4}

\usepackage{graphicx}
\usepackage{dcolumn}
\usepackage{epsfig}
\usepackage{amsmath}
\usepackage{multirow}
\usepackage{color}

\input pubboard/babarsym

\newcommand{\BABARPubYear}    {07}
\newcommand{\BABARPubNumber}  {025}
\newcommand{\SLACPubNumber} {12690}

\begin{document}
\noindent
\babar-PUB-\BABARPubYear/\BABARPubNumber\\
SLAC-PUB-\SLACPubNumber
\vskip 0.4cm

\title{\large \bf 
\boldmath
Observation of tree-level $B$ decays with $s\bar{s}$ Production from Gluon Radiation 
}

\date{\today}

%
\author{B.~Aubert}
\author{M.~Bona}
\author{D.~Boutigny}
\author{Y.~Karyotakis}
\author{J.~P.~Lees}
\author{V.~Poireau}
\author{X.~Prudent}
\author{V.~Tisserand}
\author{A.~Zghiche}
\affiliation{Laboratoire de Physique des Particules, IN2P3/CNRS et Universit\'e de Savoie, F-74941 Annecy-Le-Vieux, France }
\author{J.~Garra~Tico}
\author{E.~Grauges}
\affiliation{Universitat de Barcelona, Facultat de Fisica, Departament ECM, E-08028 Barcelona, Spain }
% ...
\collaboration{The \babar\ Collaboration}
\noaffiliation

\begin{abstract}
We report on our search for 
$\Bm \to D^{(*)+}_s K^- \pi^-$, $\Bzb \to D_s^{(*)+} \KS \pi^-$, and $\Bm 
\to D^{(*)+}_s K^- K^-$ decays in 383 million $\FourS \to B \Bbar$ events 
collected by the \babar\  detector at the \pep2\ asymmetric-energy $B$-factory. 
The decays proceed via a tree-level $b\to c$ quark transition in which a gluon 
radiates into an $s\bar{s}$ pair.
Their branching fractions are measured to be
${\cal B}(\Bm \to D^+_s K^- \pi^-)   = (2.02 \pm 0.13_{stat} \pm 0.38_{syst}) \times 10^{-4},$ 
${\cal B}(\Bm \to D^{*+}_s K^- \pi^-)= (1.67 \pm 0.16_{stat} \pm 0.35_{syst}) \times 10^{-4},$ 
${\cal B}(\Bzb \to D_s^{+} \KS \pi^-)= (0.55 \pm 0.13_{stat} \pm 0.10_{syst}) \times 10^{-4},$ and
${\cal B}(\Bm \to D_s^{+} K^- K^-)   = (0.11 \pm 0.04_{stat} \pm 0.02_{syst}) \times 10^{-4}$.
Upper limits at the 90$\%$ C.L. are set on 
${\cal B}(\Bzb \to D_s^{*+} \KS \pi^-)< 0.55 \times 10^{-4}$ and
${\cal B}(\Bm \to D_s^{*+} K^- K^-)   < 0.15 \times 10^{-4}$.
We present evidence that the invariant mass distributions of $D^{(*)+}_s K^-$ 
pairs from $\Bm \to D^{(*)+}_s K^- \pi^-$ decays are inconsistent with
the phase-space model, suggesting the presence of charm resonances lying 
below the $D^{(*)+}_s K^-$ threshold.
\end{abstract}

\maketitle

\begin{figure}[h]
\begin{center}
\includegraphics[width=0.6\linewidth]{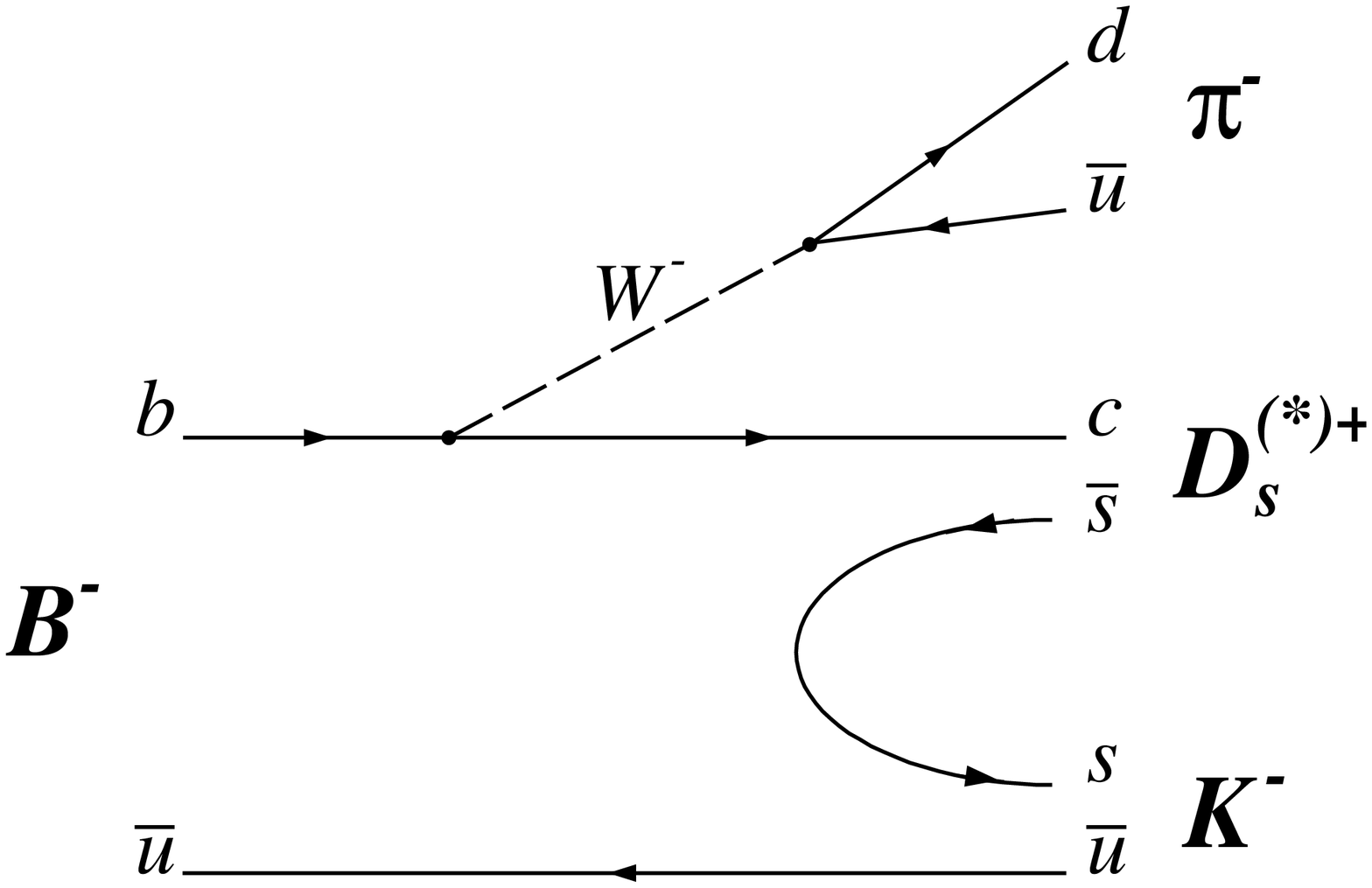}
\end{center}
\caption{Feynman diagram for $\Bm \to D_s^{(*)+} K^- \pi^-$.}
\label{fig:intro-1}
\end{figure}

Evidence for inclusive flavor correlated production
of $D_s^+$ in $\Bm$ decays was
reported recently~\cite{elba05-couderc} with a branching fraction of 
${\cal B}(\Bm \to D_s^+ X) = (1.2\pm 0.4)\%$~\cite{charge}. 
Along with $\Bm \to D_s^{*+} X$ decays, these decays are mediated by a $b\to c$ 
quark transition and require at least three final state particles,
including the production of an $s\bar{s}$ pair from the vacuum via radiative gluon pair production. 
Examples for three-body $\Bm$
decays with a  $D_s^{(*)+}$ in the final state are $\Bm \to D_s^{(*)+} K^-
\pi^-$. 
The Feynman diagram for $\Bm \to D_s^{(*)+}  K^- \pi^-$ decays are shown in
Figure~\ref{fig:intro-1}. 
The corresponding $\bar{B}^0$ decays are $\bar{B}^0\to D_s^{(*)+}
\bar{K}^0 \pi^-$. By replacing the $\pi^-$ in Figure~\ref{fig:intro-1} with a $K^-$, we get the
Cabibbo-suppressed decays $\Bm\to D_s^{(*)+} K^- K^-$.
 
In addition to the dominant
diagram, $\Bm \to D_s^{(*)+} K^- \pi^-$ can occur via the 
color-suppressed diagram where the constituent $\bar{u}$'s of the $K^-$ and $\pi^-$ are switched.
Although a color-suppressed contribution does not exist for 
$\bar{B}^0 \to D_s^{(*)+} \bar{K}^0 \pi^-$, 
a sub-dominant contribution from a $W$-exchange diagram
with $s{\bar s}$ and $d{\bar d}$ popping may exist instead.
Either of these contributions could cause a deviation
from the naive expectation of two for the ratio of 
$\Bm \to D_s^{(*)+} K^- \pi^-$ to $\Bzb \to D_s^{(*)+} \KS \pi^-$ branching fractions.

The $D_s^{(*)}$-$K$ pair could come from intermediate charm resonances
instead of directly from the $B$.
It has been proposed that these resonances can play a significant role in $\Bm \to D^{+}_s K^- \pi^-$
decays~\cite{theo-paper} despite their masses lying below the $m(D_s K)$ production 
threshold~\cite{belle-dpipi}. 
In this case, it may be possible to measure the parameters
of the resonances 
such as their masses and widths, complementary to the analysis
using $B \to {\bar D} \pi \pi$ decays~\cite{belle-dpipi}.

Along with exclusive $\Bm \to D_s^{(*)+}X$ and $\bar{B}^0 \to D_s^{(*)+}X$ three-body decays, 
no decays proceeding via radiative gluon $s\bar{s}$ pair production at the tree level
have hitherto been observed.
Upper limits on the branching fractions of the $\Bm\to D_s^{(*)+}  K^-
\pi^-$ and $\Bzb \to D_s^{(*)+} \KS \pi^-$ modes have been placed by ARGUS~\cite{argus-limit}.
In this paper we report first observations of the decay modes $\Bm\to D_s^{(*)+}  K^- \pi^-$
and  $\Bzb \to D_s^{+} \KS \pi^-$, evidence for $\Bm \to D^+_s K^- K^-$, 
and limits on the branching fractions of 
$\Bzb \to D_s^{*+} \KS \pi^-$ and $\Bm \to D^{*+}_s K^- K^-$.  
We also present a measurement of an invariant mass distribution of $D_s^{(*)+} K^-$ 
pairs from $\Bm \to D_s^{(*)+} K^- \pi^-$ decays and compare it to the spectrum
obtained from a phase-space model.

The analysis uses approximately 383 million $\FourS \to B \Bbar$ events created by the \pep2\
$e^+ e^-$ collider and collected by the \babar\ detector. 
The \babar\ detector is described elsewhere~\cite{refbabar}.

Optimal selection criteria and probability density functions of selection
variables are determined by an analysis based on Monte Carlo (MC)
simulation of both signal and background events. 
We use GEANT4~\cite{geant} software to
simulate interactions of particles traversing the \babar\ detector,
taking into account the varying detector conditions and beam
backgrounds. We verify with MC simulation that resolutions and
background levels correctly describe the data.  

\renewcommand{\multirowsetup}{\centering}
\newlength{\LL}\settowidth{\LL}{$8047$}
\begin{table*}[!htb]
\begin{center}
\caption{Summary of results for the total detection efficiencies
$\eps$ excluding the subsequent branching fractions of $D_s^{(*)}$ decay modes 
($D_s^{*+} \ra D_s^+ \gamma,~D_s^+ \ra \phi \pi^+,~\Kstarzb K^+,~\KS K^+$), 
expected peaking background events $n_{\rm peaking}$ with statistical uncertainties from fits
of the $m_{ES}$ distributions obtained 
using 
the $D^+_s$ invariant mass sidebands, 
final signal ($n_{sig}$) and background ($n_{bkg}$) yields with statistical uncertainties from 
$m_{ES}$ fits adjusted to account for 
estimated peaking 
backgrounds, and cross-feed 
contributions,  
branching fractions ${\cal B}$ with statistical and systematic uncertainties,
significances $s(\sigma)$ calculated by comparing the likelihood maximum 
of the nominal fit to that of the fit with 
the signal yield fixed to the difference between the raw and corrected signal yields,
and upper limits UL on the branching fractions for 
$\Bzb \to D_s^{*+} \KS \pi^-$ and $\Bm \to D^{*+}_s K^- K^-$. The background yields
$n_{bkg}$ are calculated in the region $5.27-5.29$~\gevcc\ .} 
\label{tab:summary}
\begin{ruledtabular}
\begin{tabular}{lccccccccc}
Mode &    
$\eps_{\phi \pi}$& $\eps_{\bar{K^*} K}$& $\eps_{\KS K}$&
$n_{\rm peaking}$&
$n_{sig}$ & $n_{bkg}$ & ${\cal B} \times 10^{-4}$& 
$s(\sigma)$ &
UL (\@90$\%$ C.L.) \\[1mm]
\hline\\[-2mm]
$\Bm \ra D^+_s K^- \pi^-$ &    11.1$\%$& 6.8$\%$& 9.6$\%$& $41 \pm 9$ &    $430 \pm 29$   & $182 \pm 6$   & $2.02 \pm 0.13 \pm 0.38$ & 21 & ---  \\ [2mm]
                                                                                                                                  
$\Bm \ra D^{*+}_s K^- \pi^-$&  5.9$\%$& 3.6$\%$& 5.1$\%$ & $4 \pm 5$  &    $178 \pm 18$   & $87.1 \pm 3.5$& $1.67 \pm 0.16 \pm 0.35$ & 14 & ---  \\ [2mm]
                                                                                                                            
$\Bzb \to D_s^+ \KS \pi^-$  &  8.8$\%$& 5.3$\%$& 7.6$\%$ & $28 \pm 6$  &   $61.8 \pm 14.4$& $94.5 \pm 5.5$& $0.55 \pm 0.13 \pm 0.10$ & 5.2& ---  \\ [2mm]
                                                                                                                            
$\Bzb \to D_s^{*+} \KS \pi^-$& 3.8$\%$& 2.3$\%$& 3.4$\%$ & $-1.1 \pm 2.7$& $13.6 \pm 8.4$ & $62.8 \pm 3.4$& $0.29 \pm 0.18 \pm 0.073$& 1.8& $0.55 \times 10^{-4}$\\[2mm]
                                                                                                                            
$\Bm \to D_s^{+} K^- K^-$&     7.1$\%$& 4.3$\%$& 6.3$\%$&  $-0.3 \pm 1.9$& $14.4 \pm 5.6$ & $9.8 \pm 1.3$ & $0.11 \pm 0.04 \pm 0.02$ & 3.3& ---  \\ [2mm]
                                                                                                                            
$\Bm \to D_s^{*+} K^- K^-$ &   3.8$\%$& 2.4$\%$& 3.5$\%$ & $-1.7 \pm 1.3$& $4.7 \pm 4.0$  & $6.5 \pm 0.9$ & $0.07 \pm 0.06 \pm 0.02$ & 1.3& $0.15 \times 10^{-4}$\\
\end{tabular}
\end{ruledtabular}
\end{center}
\end{table*}

Candidates for \Ds mesons are reconstructed in the modes $\Ds\to
\phi\pi^+$, $\Kstarzb K^+$, and $\KS K^+$, with $\phi\to K^+K^-$,
$\Kstarzb\to K^-\pi^+$ and $\KS\to\pi^+\pi^-$. 
The \KS candidates
are reconstructed from two oppositely-charged tracks 
that come from a
common vertex displaced from the $e^+e^-$ interaction point.
We require the significance of this displacement (the measured $\KS$ flight distance
divided by its estimated error) to exceed 2.
All other tracks are required to originate less than 1.5~cm away from the
$e^+e^-$ interaction point in the transverse plane and less than 10~cm
along the beam axis.
Charged kaon candidates must satisfy identification criteria
that are typically around 92\% efficient~\cite{kaon}, depending on momentum and
polar angle, and have a pion misidentification rate at the 5$\%$ level.
The $\phi\to K^+K^-$, 
$\Kstarzb\to K^-\pi^+$ and $\KS\to\pi^+\pi^-$ candidates are required to have invariant
masses within $\pm 15~\mevcc$, $\pm 50~\mevcc$ and $\pm 10~\mevcc$ of their nominal masses, 
respectively~\cite{pdg}.
The full polarization of the \Kstarzb and $\phi$ mesons from the \Ds decays are
exploited to reject backgrounds through the use of the helicity
angle $\theta_H$, defined as the angle between the $K^-$ momentum vector and the
direction of flight of the \Ds in the 
\Kstarzb or $\phi$ rest frame. 
The \Kstarzb and $\phi$ candidates are required to have $|\cos\theta_H|$ greater than 0.5. 

The $D^{*+}_s$ candidates are reconstructed in the mode $D^{*+}_s \ra D^+_s \gamma$.
The photons are accepted if their energy is greater than 100~\mev.
Photons from $D^{*+}_s$ candidates are rejected if, when combined with any other photon having an energy
greater than 150~\mev, they belong to a photon pair whose invariant mass lies within $\pm 10~\mevcc$ 
of the $\pi^0$ mass.  
The \Ds candidates are
required to have invariant masses in the interval $\pm 10$~\mevcc of the nominal \Ds mass 
while the invariant masses of $D^{*+}_s$ candidates lie in the range 
from $m(D^{*+}_s)-15~\mevcc$ to $m(D^{*+}_s)+10~\mevcc$.
All $D^+_s$ candidates are subjected to a mass-constrained fit after selection.
The invariant mass of the
$D^{*+}_s$ is calculated after the mass constraint on the daughter $D^+_s$ has been applied.
Subsequently, all $D^{*+}_s$ candidates are subjected to mass-constrained fits. 
To eliminate $\Bzb \to D_s^{(*)+} D^-,~D^- \to \KS \pi^-$ events from the 
$\Bzb \to D_s^{(*)+} \KS \pi^-$ samples, the invariant mass of the $\KS$ and $\pi^-$ 
must be outside a 40~\mevcc window around the $D^-$ mass. 

Finally, the $B$ meson candidates are formed using the reconstructed
combinations of $D^+_s K^- \pi^-$, $D^{*+}_s K^- \pi^-$, $D_s^{+} \KS \pi^-$, 
$D_s^{*+} \KS \pi^-$, $D_s^{+} K^- K^-$, and $D_s^{*+} K^- K^-$.

The background from continuum $q\bar{q}$ production (where $q = u,d,s,c$) is
suppressed based on the event topology. 
The event shape variables, $R_2$ (the ratio of the second to zeroth Fox-Wolfram
moments~\cite{fox-wolfram}) and $L_2/L_0$ (the ratio of the second and zeroth angular moments of the energy flow
about the $B$ thrust axis~\cite{l-moments}), are combined in a Fisher discriminant ($\cal{F}$) 
to effectively exploit the difference between the shapes of $e^+ e^- \ra B \bar{B}$ 
and $e^+ e^- \ra q \bar{q}$ events. 
A selection is applied to ${\cal F}$ such that 80$\%$ of continuum
background is rejected while maintaining 80$\%$ signal efficiency.

The signals are extracted using the energy-substituted mass
$\mes \equiv \sqrt{E_{\rm b}^{*2} - (\sum_i {\mathbf p}^*_i)^2}$ and the energy difference
$\Delta E \equiv {\sum_i}{\sqrt{m_i^2+{\mathbf p}_i^{*2}}} - E_{\rm b}^*$, 
where $E_{\rm b}^*$ is
the beam energy in the laboratory frame, 
${\mathbf p}^*_i$ is the momentum of the daughter particle $i$ of the 
$B$ meson candidate also in the laboratory frame, and
$m_i$ is the mass hypothesis for particle $i$. For signal events, \mes
peaks at the $B$ meson mass with a resolution of about 2.6~\mevcc and
$\Delta E$ peaks near zero with a resolution of 13~MeV.
The $B$ candidates are required to have $|\Delta E|<
25\ \mev$ and $\mes > 5.2~\gevcc$.
After all selection criteria are applied, we find the fraction of events containing more than one $B$
candidate to be between 3\% and 11\% depending on the decay mode. In these instances, the $B$ candidate with $\Delta E$ closest 
to zero is chosen. The estimated $B$
reconstruction efficiencies, excluding the subsequent branching fractions, are shown in 
Table~\ref{tab:summary}.

Background events that pass these selection criteria are represented by
approximately equal amounts of $q\bar{q}$ continuum and $B{\bar B}$ events. We parametrize
their \mes distributions by a
threshold function~\cite{Argus}: 
$$
f(m_{\rm ES}) \sim m_{\rm ES} \sqrt{1-x^2} {\rm exp}[-\xi (1-x^2)],
\label{eq:argus}
$$
where $x = 2 m_{\rm ES}/\sqrt{s}$, $\sqrt{s}$ is the total energy
of the beams in their center of mass frame, and $\xi$ is a
fit parameter. 

A study using simulated $B$ decays reveals significant numbers of
background events peaking in the regions of  $5.272 < \mes < 5.288~\gevcc$ and $|\Delta E|< 25~$MeV
similar to the reconstructed signal candidates. 
This peaking background
is due to charmless and charmonium
$B$ decays with the same set of particles as signal in the final state.
The peaking contribution is evaluated using the data
by reconstructing $D^{(*)+}_s K^- \pi^-$, $D^{(*)+}_s \KS \pi^-$ and $D^{(*)+}_s K^- K^-$  
combinations, where
``${D^+_s}$'' candidates are selected from 25 - 40 \mevcc sidebands
around the $D^+_s$ nominal mass.
In this procedure, we use the same selection requirements as for the
signal except that ``$D^+_s$'' candidates are not mass
constrained. Studies revealed that constraining the $D^+_s$ mass did not
significantly affect the resolutions of \mes\ and \DeltaE distributions
and that events in the $D^+_s$ mass sidebands are a good representation
of the background under the $D^+_s$ peak. Table~\ref{tab:summary} shows
the fit yields of the peaking  background contribution under the \mes
peak for each mode.

A matrix is constructed to study the cross-feed between the
signal modes. Its elements describe 
the contributions of each mode according to the levels seen in MC samples.
No off-diagonal element of the cross-feed matrix exceeds $2\%$;
this near-diagonal structure indicates effective suppression of the cross-feed 
contributions by application of the selection criteria.

\begin{figure}[!tbp]
\begin{center}
\includegraphics[width=0.50\linewidth]{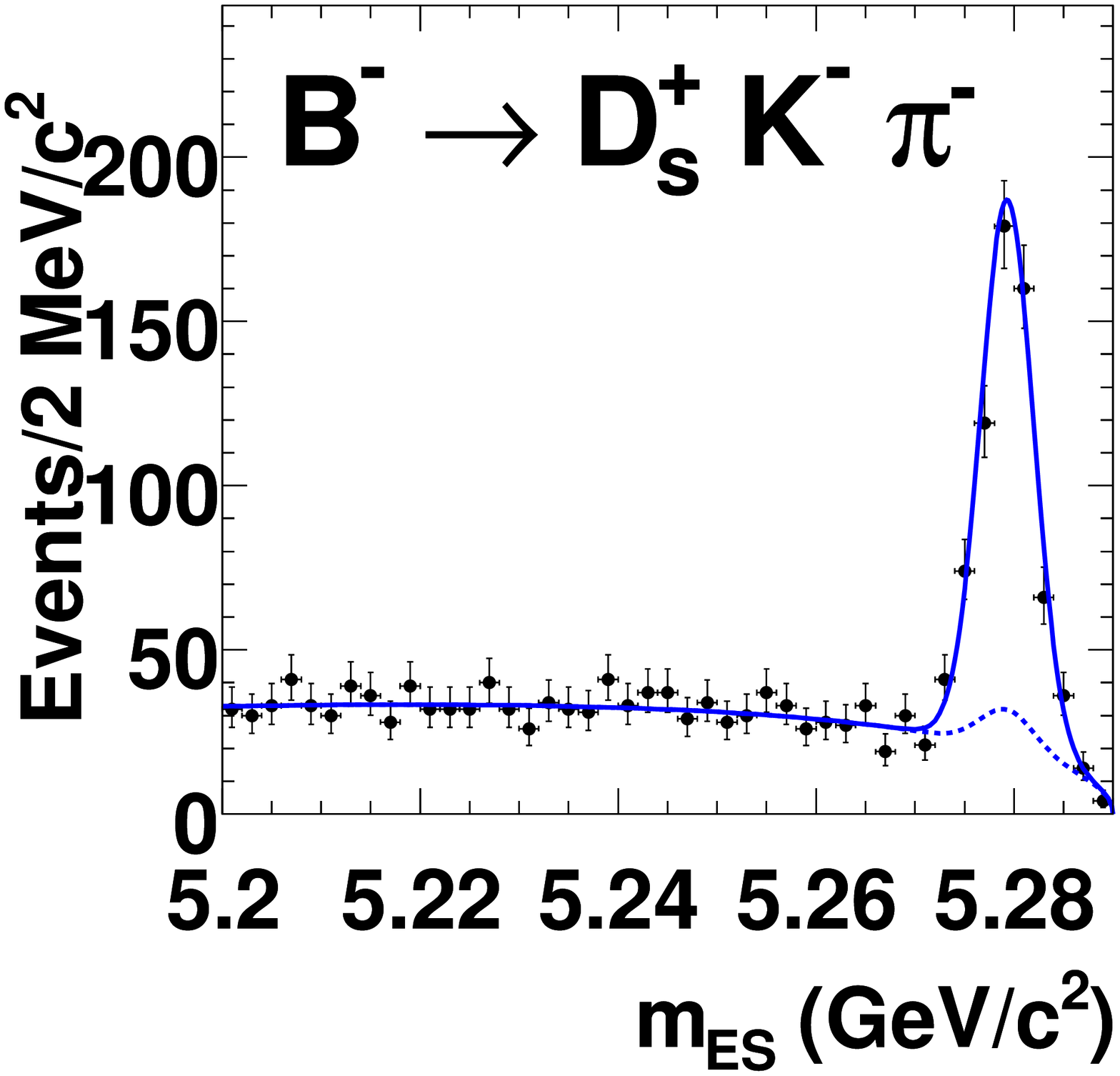}\includegraphics[width=0.50\linewidth]{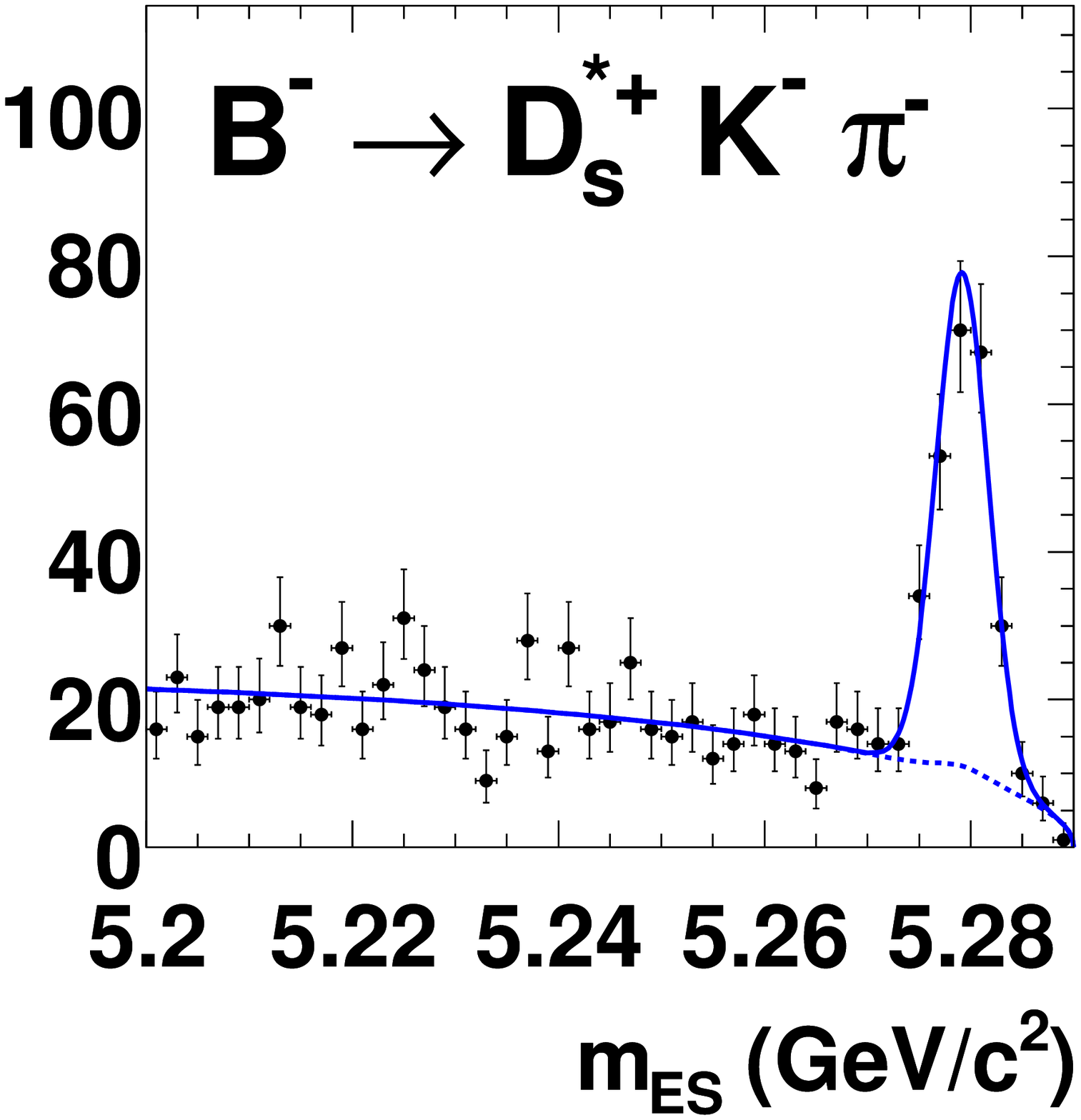}
\vskip-9mm
\includegraphics[width=0.50\linewidth]{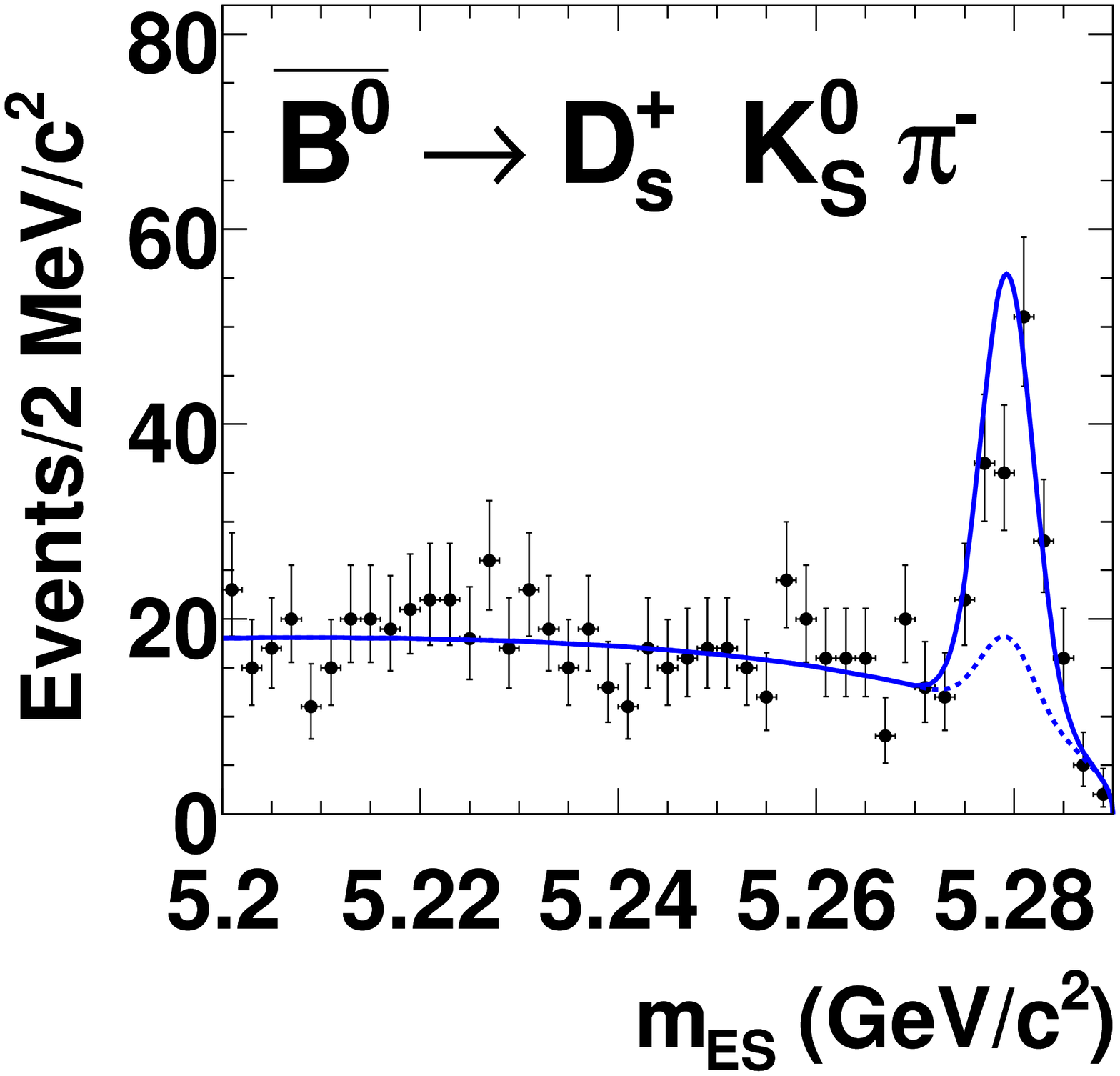}\includegraphics[width=0.50\linewidth]{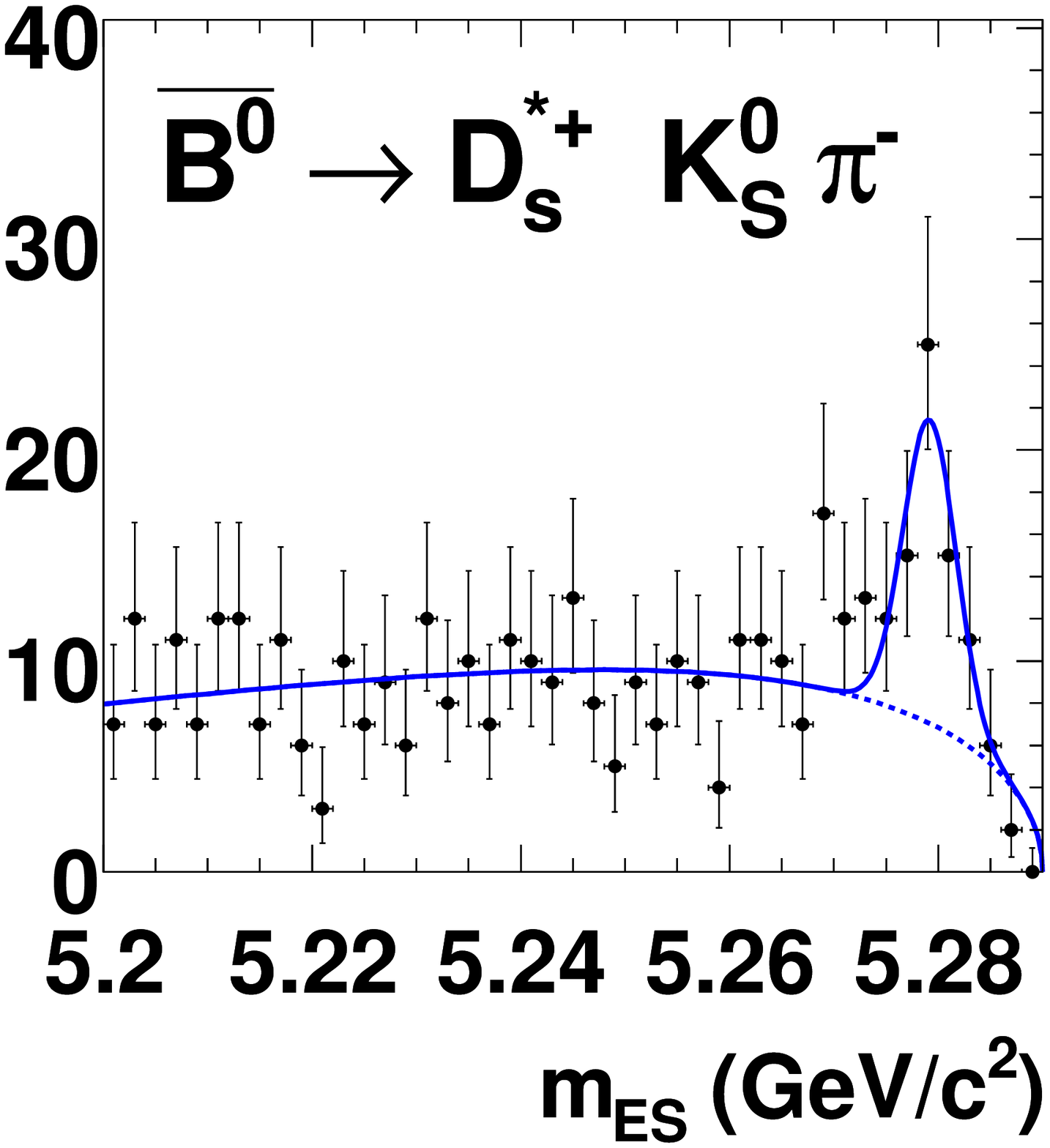}
\vskip-9mm
\includegraphics[width=0.50\linewidth]{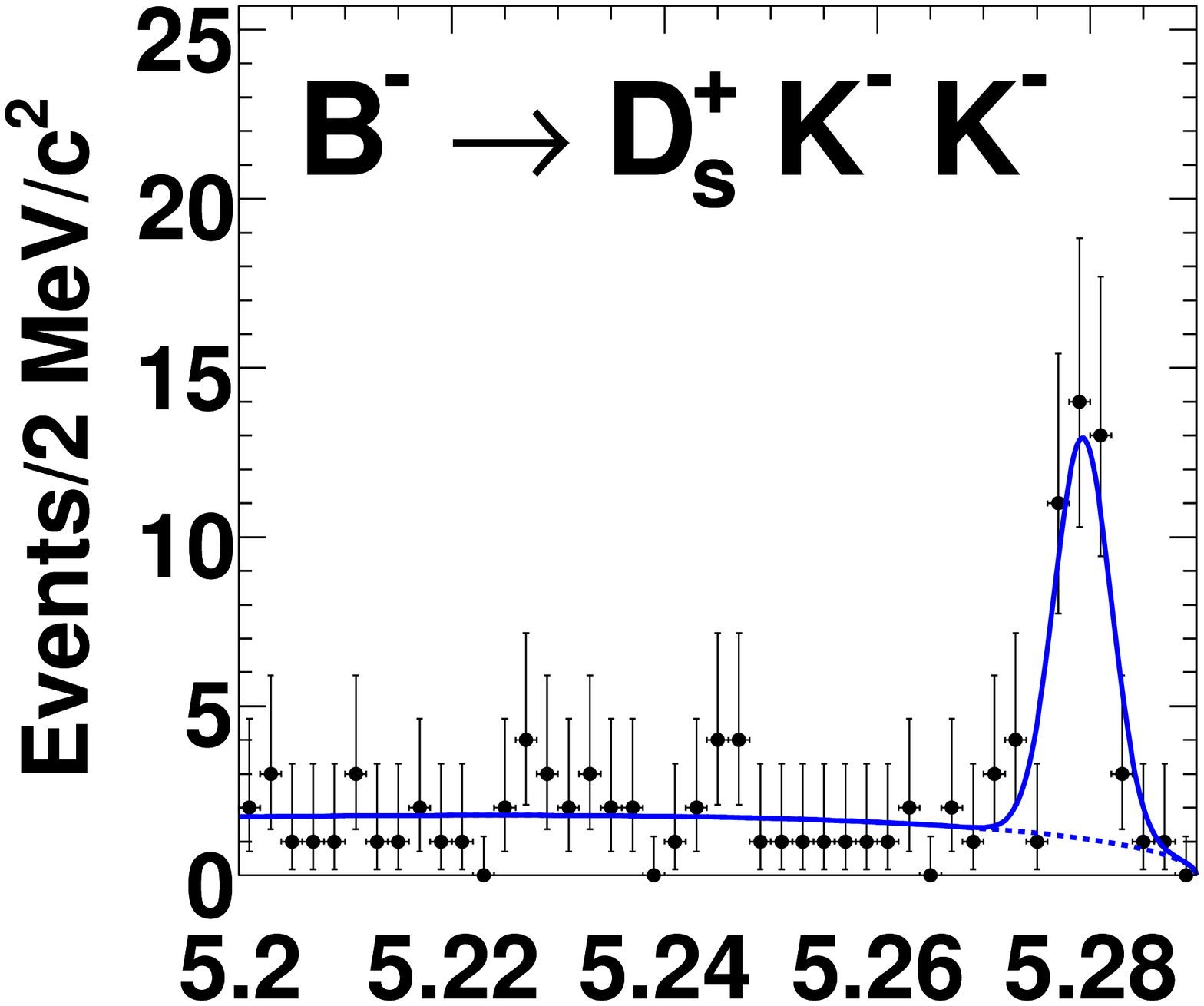}\includegraphics[width=0.50\linewidth]{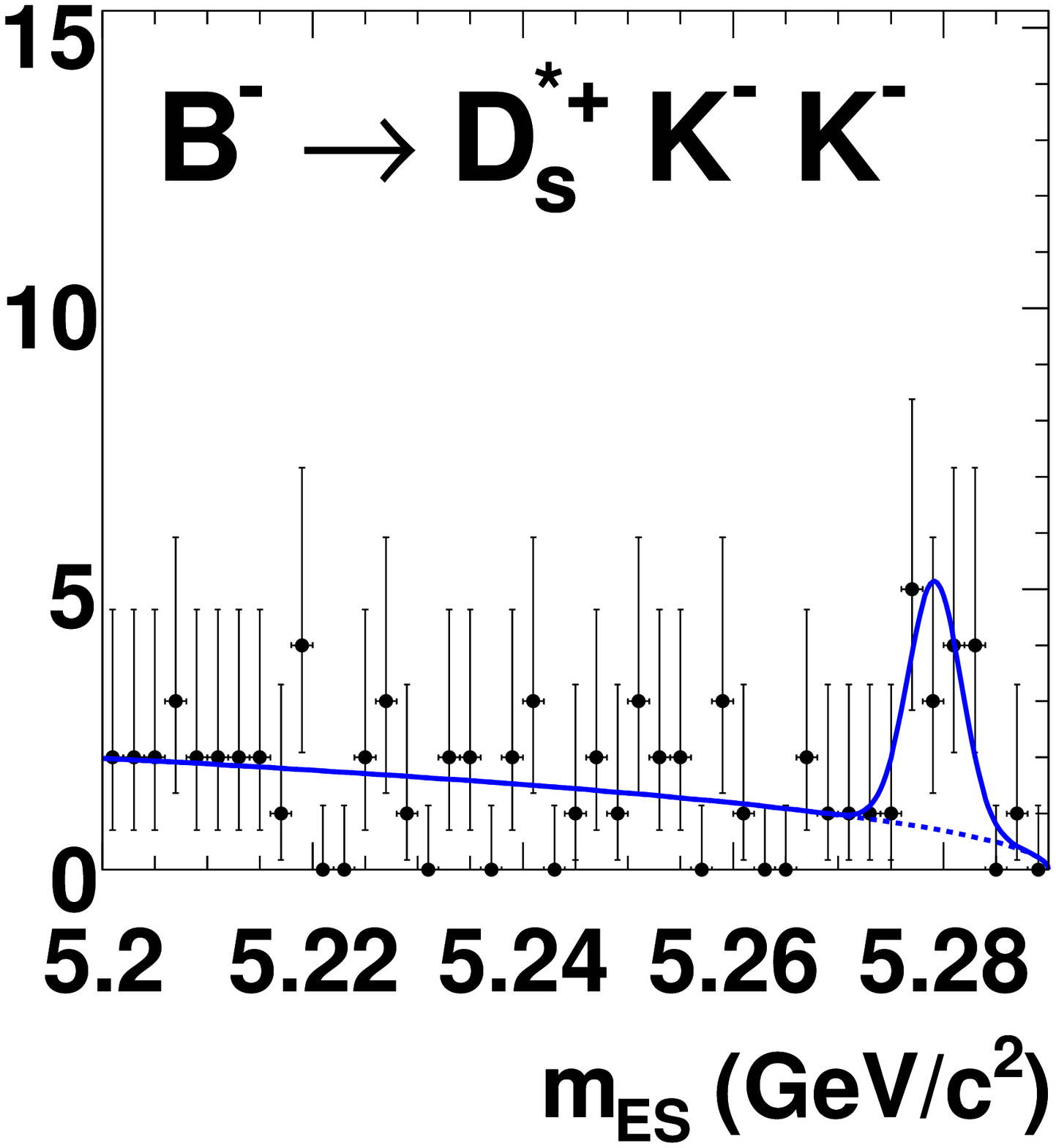}
\caption{\label{fig:mes-all} $m_{ES}$ spectra for the $\Bm \to D^+_s K^- \pi^-$ (top left), 
$\Bm \to D^{*+}_s K^- \pi^-$ (top right), $\Bzb \to D_s^{+} \KS \pi^-$ (middle left), 
$\Bzb \to D_s^{*+} \KS \pi^-$ (middle right), $\Bm \to D_s^{+} K^- K^-$ (bottom left), 
and $\Bm \to D_s^{*+} K^- K^-$ (bottom right). Solid curves
show the fit results, as explained in the text. Dashed lines in the signal
regions correspond to the peaking and non-peaking background components of the fit. 
The data are the points with error bars.}
\end{center}
\end{figure}

Figure~\ref{fig:mes-all} shows the \mes
spectra of the reconstructed $B$ candidates.
For each mode, we perform an
extended unbinned maximum likelihood (ML) fit to the \mes distributions using the
candidates from all \Ds decay modes combined. The \mes
distributions are fit with the 
sum of two functions: $f(m_{\rm ES})$ characterizing
the combinatorial background and a Gaussian function to describe the
signal.
The likelihood function is given by:  
$$
{\cal L} = \frac{e^{-(n_{sig}+n_{bkg})}}{N!} \prod_{i=1}^N (n_{sig} P_i^{sig} + n_{bkg} P_i^{bkg}),
\label{eq:likelihood}
$$
where 
$P^{sig}_i$ and $P^{bkg}_i$ are the probability density functions
for the signal and background, $n_{sig}$ and $n_{bkg}$ are the number of signal and background 
events, and $N$ is the total number of events in the fit

The final signal yields are obtained by subtracting the estimated peaking background 
and cross-feed contributions from the yields of the $m_{ES}$ fits described in the preceding 
paragraph.
No peaking background is subtracted from modes that have $n_{\rm peaking}$ less than 
zero in
Table~\ref{tab:summary} since these values are consistent with zero although their 
errors are still propagated.
The final values are given in the $n_{sig}$ column of Table~\ref{tab:summary}. 
The total signal yield in each $B$ decay mode is related to the $B$ branching fraction 
${\cal B}$ using the following
expression:
$$
{\cal B} = n_{sig}/(N_{B \bar B} \cdot \sum_i  {\cal B}_i \cdot \eps_i),
$$
where $N_{B \bar B}$ is the number of produced $B\bar B$ pairs,
${\cal B}_i$ is the product of the intermediate branching ratios,
$\eps_i$ is the reconstruction efficiency (from Table~\ref{tab:summary}) 
and the sum is over $D^+_s$ modes ($i=\phi \pip$, $\Kstarzb K^+$, $\KS K^+$).
As an input to the calculations, we used branching fraction numbers
from~\cite{pdg}. 
The results of these calculations are summarized in Table~\ref{tab:summary}.

The total relative systematic uncertainty in the $B$ branching fractions
is estimated to be approximately $19\% - 25\%$ depending on the decay
mode.
The largest contribution, an uncertainty of $15\%$, comes from the 
$D^+_s$ branching fractions. 
The differences between selection 
efficiencies in MC and in the data (estimated using the control mode
$\Bm \to D^-_s D^0,~D^0 \to K^- \pi^+$) contribute to the systematic uncertainty 
($5\% - 10\%$) as does the efficiency dependence on the $D^{(*)+}_s K^-$
invariant mass spectrum ($7\% - 9\%$). 
In the \mes\ fits of the lower statistics modes ($D^{*+}_s \KS \pi^-$
and $D^{*+}_s K^- K^-$) the signal Gaussian 
parameters and $\sqrt{s}$ in $f(m_{\rm ES})$ are fixed to ensure fit convergence.
The associated systematic uncertainties are $14\%$ and $9\%$, respectively.
The entries in the cross-feed matrix affecting the $D^{(*)+}_s K^- K^-$ modes
vary by $8\%(5\%)$ when they are calculated 
with MC events weighted according to the observed spectra of the 
$D^{(*)+}_s K^-$ invariant mass.

For the $\Bzb \to D_s^{*+} \KS \pi^-$ and $\Bm \ra D^{*+}_s K^- K^-$
decay modes, the upper limits are set using a frequentist
approach~\cite{pdg} and taking into account the systematic
uncertainties. The upper limits are summarized in 
Table~\ref{tab:summary}. 

\begin{figure}[ht!]
\begin{center}
\includegraphics[width=4.3cm]{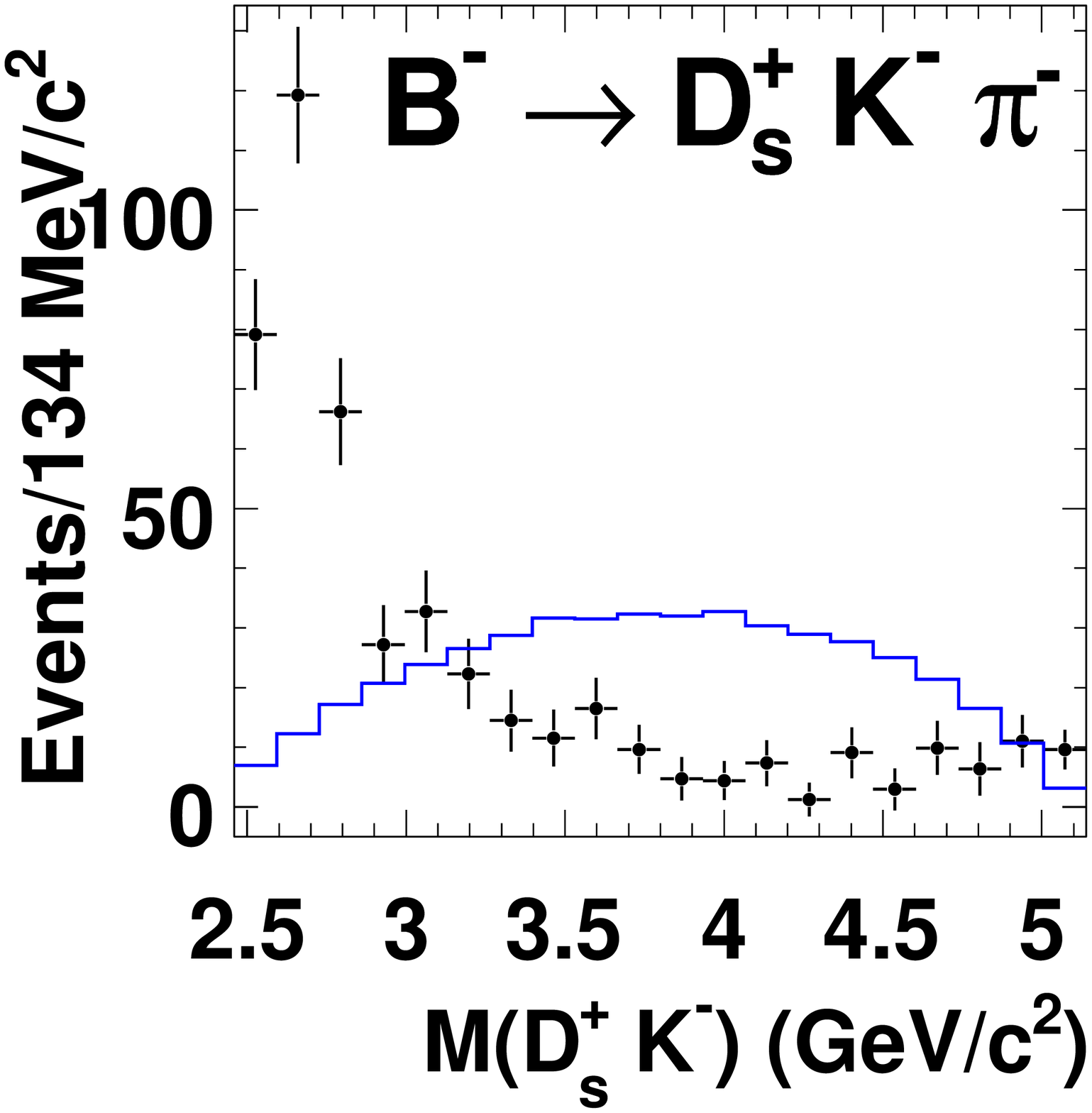}\includegraphics[width=4.3cm]{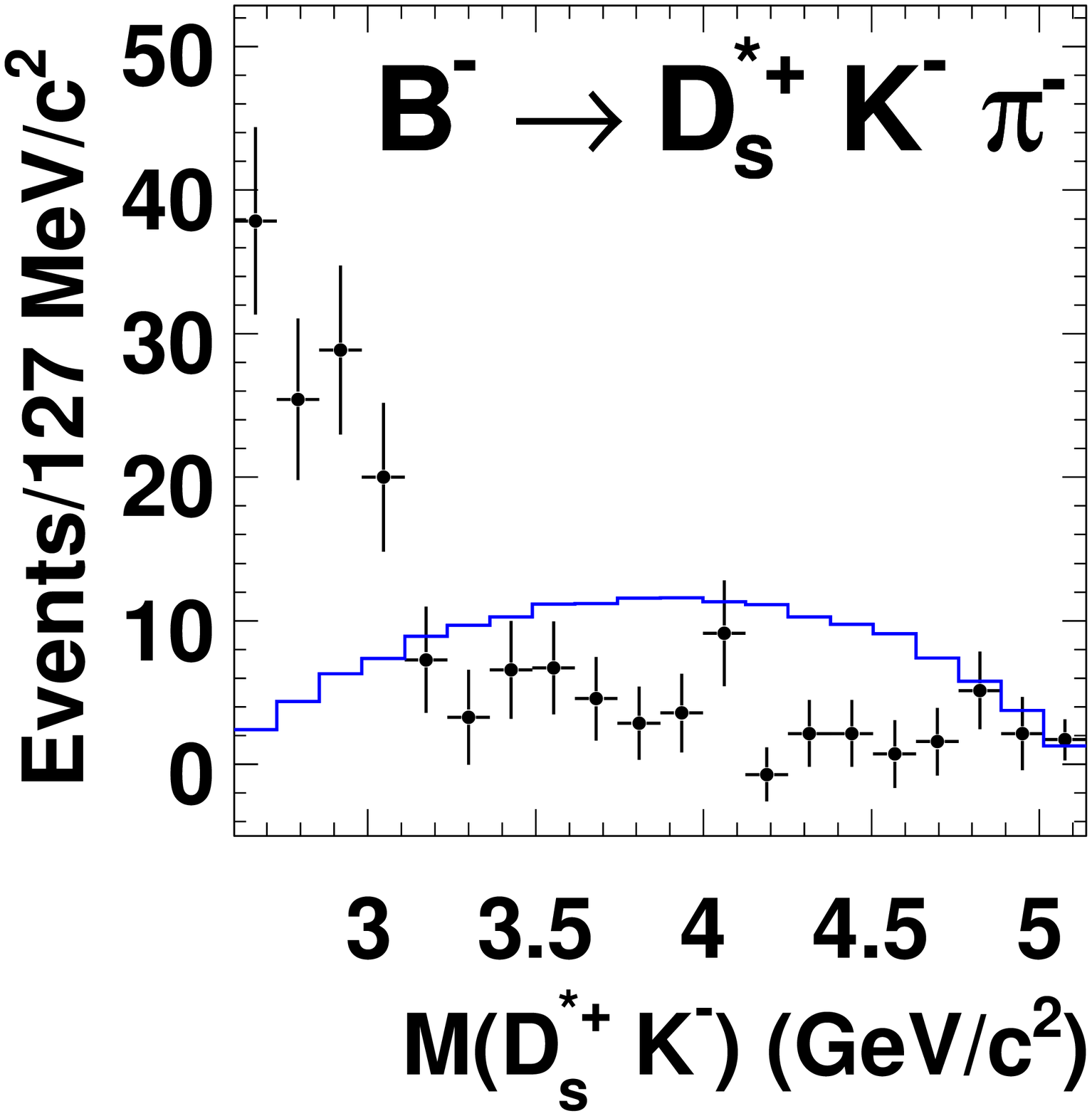}
\caption{\label{fig:unblinded-dsk} $D^{(*)+}_s K^-$ invariant mass spectra for the 
$\Bm \to D^+_s K^- \pi^-$ (left)
and $\Bm \to D^{*+}_s K^- \pi^-$ (right) decay modes using the data. A requirement of 
$m_{ES}>5.270$~\gevcc is applied to the events shown in the figure, in 
addition to the signal selection described in the text. 
Combinatoric background is approximated and then subtracted using events outside the \mes signal region
($m_{ES}<5.265$~\gevcc). 
The histogram shows the non-resonant
signal MC events distribution, scaled to the number of events in the data signal region.}
\end{center}
\end{figure}

Studies of the invariant mass spectra of the $D_s^{(*)+} K^-$ system in
$\Bm \to D^{(*)+}_s K^- \pi^-$ modes reveal distributions incompatible 
with those of three-body phase space. As shown in Figure~\ref{fig:unblinded-dsk}, 
there are enhancements in the number of events at the lower ends of the $m(D_s^{(*)+} K^-)$ 
spectra. These enhancements suggest the presence of charm resonances 
lying below the $D^{(*)+}_s K^-$ threshold~\cite{theo-paper}. 

In summary, 
$\Bm \to D^+_s K^- \pi^-$, $\Bm \to D^{*+}_s K^- \pi^-$ and
$\Bzb \to D_s^{+} \KS \pi^-$ decays are observed
for the first time each with significance greater than $5 \sigma$. 
Evidence for $\Bm \to D_s^{+} K^- K^-$ was found with a significance slightly greater than $3 \sigma$. 
Upper limits are set on the branching fractions of the two decay modes with significances lower than
 $2 \sigma$:
$\Bzb \to D_s^{*+} \KS \pi^-$ and $\Bm\to D_s^{*+} K^- K^-$. 

The ratios of ${\cal B}(\Bm \to D_s^{(*)+} K^- K^-)$ to ${\cal B}(\Bm \to D^{(*)+}_s K^- \pi^-)$ are 
consistent with the expected Cabibbo suppression. The branching fraction of 
$\Bzb \to D_s^{+} \KS \pi^-$ is less than half that of $\Bm \to D^+_s K^- \pi^-$; 
this may be due to the W-exchange diagram correction to the neutral mode and
the color-suppressed contribution to the charged mode.

We are grateful for the excellent luminosity and machine conditions
provided by our \pep2\ colleagues, 
and for the substantial dedicated effort from
the computing organizations that support \babar.
The collaborating institutions wish to thank 
SLAC for its support and kind hospitality. 
This work is supported by
DOE
and NSF (USA),
NSERC (Canada),
CEA and
CNRS-IN2P3
(France),
BMBF and DFG
(Germany),
INFN (Italy),
FOM (The Netherlands),
NFR (Norway),
MIST (Russia),
MEC (Spain), and
STFC (United Kingdom). 
Individuals have received support from the
Marie Curie EIF (European Union) and
the A.~P.~Sloan Foundation.


\begin{thebibliography}{99}
\bibitem{charge}
Charge conjugate reactions are implicitly included, throughout this paper.
\relax
\end{thebibliography}
\end{document}